\pdfoutput=1
\documentclass[twocolumn,superscriptaddress,aps,preprintnumbers,amsmath,amssymb,prd,nofootinbib]{revtex4-2}

\usepackage{amsmath}
\usepackage{amssymb}
\usepackage{amsfonts}
\usepackage{graphicx}
\usepackage{xcolor}
\usepackage{xfrac}
\usepackage{comment}
\usepackage{pifont}
\usepackage{physics}
\usepackage{fourier}
\usepackage{hyperref}
\usepackage{bm}
\usepackage{enumitem}

\definecolor{rossoferrari}{HTML}{D9073D}
\definecolor{mediumblue}{HTML}{0000CD}
\definecolor{forestgreen}{HTML}{228B22}
\definecolor{desy_blue}{HTML}{009EE2}
\definecolor{desy_orange}{HTML}{FD8800}
\definecolor{light_pink}{rgb}{1,0.4,0.4}
\definecolor{light_blue}{rgb}{0.284602,0.317763,0.963947}
\hypersetup{setpagesize=false,bookmarksnumbered=true,bookmarksopen=true, colorlinks=true,linkcolor=light_blue,urlcolor=rossoferrari,citecolor=rossoferrari,linktocpage=false}

\usepackage{slashed}

\renewcommand{\thefootnote}{\fnsymbol{footnote}}

\newcommand{\bea}{\begin{array}}
\newcommand{\eea}{\end{array}}
\newcommand{\beq}{\begin{eqnarray}}
\newcommand{\eeq}{\end{eqnarray}}

\newcommand{\lmk}{\left(}  
\newcommand{\rmk}{\right)}

\def\eq#1{Eq.~(\ref{#1})}

\definecolor{orange}{RGB}{255,100,0}
\definecolor{rosepink}{RGB}{248,100,100}

\begin{document}

\title{A Small-Throat Boundary Condition for the Tunneling Wave Function of the Universe}

\author{Masaki Yamada}
\email{m.yamada@tohoku.ac.jp}
\affiliation{Department of Physics, Tohoku University, Sendai, Miyagi 980-8578, Japan}

\preprint{TU-1311}

\date{\today}


\begin{abstract}
We propose a small-throat prescription for the wave function of a closed universe in the semiclassical approximation, motivated by the idea that universe creation may be obtained as the decoupling, or pinch-off, limit of a tunneling geometry connected to another universe through a small throat.  Instead of retaining the parent-universe side explicitly, we describe the remaining half-geometry by a minisuperspace model with boundary conditions imposed at the throat.
To model the finite throat, we introduce a small radiation component parametrized by $\epsilon$ in a closed minisuperspace model with a positive cosmological constant.
The radiation term produces two turning points, an inner one $q_-\sim O(\epsilon)$ and an outer one $q_+\sim H^{-2}$, where $q$ is the square of the scale factor. 
Our prescription imposes the Neumann condition $\dot q(0)=0$ on the classical solutions and restricts the initial size $q_i=q(0)$ to the small-throat domain $0<|q_i|<\sqrt{\epsilon}/H$. This domain contains the inner turning point $q_-$ and excludes the unsuppressed outer-turning-point saddle at $q_i=q_+$. We choose the steepest-descent lapse contour through the inner tunneling saddle in the semiclassical approximation.
Taking the limit $\epsilon\to0$ after this finite-throat saddle problem has been defined, the inner turning point approaches $q_i=0$, and the saddle action reduces to that of the standard tunneling saddle.  Other choices of lapse contour can instead select Hartle--Hawking-type growing branches.  In this sense, the tunneling exponent and outgoing WKB phase can be obtained as the limiting form of tunneling from an arbitrarily small universe, rather than by imposing a boundary condition directly at a vanishing geometry.  
\end{abstract}

\maketitle

\renewcommand{\thefootnote}{\arabic{footnote}}
\setcounter{footnote}{0}

\section{Introduction}

Understanding the quantum origin of the universe is a central problem in
quantum cosmology.  Although a complete answer may ultimately require a full
theory of quantum gravity, minisuperspace models provide a useful arena in
which the wave function of the universe and the corresponding gravitational
path integral can be studied explicitly.  In the canonical language this
problem is often phrased in terms of solutions to the Wheeler--DeWitt equation with a chosen boundary condition,
while in the path-integral language one must specify both a boundary condition and an
integration contour.  Representative proposals include the tunneling wave
function and the Hartle--Hawking no-boundary wave function
\cite{DeWitt:1967yk,Hartle:1983ai,Vilenkin:1983xq,Vilenkin:1984wp,
Linde:1983mx,Rubakov:1984bh,Zeldovich:1984vk,Vilenkin:1986cy,
Vilenkin:1987kf,Vilenkin:1994rn}.  Their path-integral realizations and
minisuperspace saddle-point structures have been studied in many works,
including Refs.~\cite{Louko:1987wq,Halliwell:1988ik,Halliwell:1989dy,
Brown:1990iv,Feldbrugge:2017kzv}.

A basic difficulty of Euclidean gravitational path integrals is the
conformal-factor problem: the Euclidean Einstein action is not bounded from
below \cite{Gibbons:1978ac}.  Lorentzian formulations avoid this particular
unboundedness, but the path integral is then oscillatory and requires a
prescription for convergence and for the selection of saddle points.  A
powerful way to address this issue is to complexify the lapse and use
Picard--Lefschetz theory, in which the original lapse contour is deformed into
steepest-descent contours attached to relevant saddles
\cite{Halliwell:1988ik,Halliwell:1989dy,Feldbrugge:2017kzv}.  In the
closed de Sitter minisuperspace model, this Lorentzian analysis selects the
tunneling saddle for the standard real-lapse contour, and recent thimble and
resurgence analyses further clarify the associated contour structure
\cite{Honda:2024aro,Chou:2024sgk}.%
\footnote{
There are also proposals to reproduce the Hartle--Hawking wave function by
choosing intrinsically complex lapse contours
\cite{DiazDorronsoro:2017hti,DiazDorronsoro:2018wro,Janssen:2019sex}.
Although 
we focus on the contour and saddle structure relevant to the
tunneling branch in this paper, the Hartle--Hawking wave function can be reproduced by choosing a different integration contour.
}
In these prescriptions, the path integral is defined with a vanishing geometry at the initial boundary.

\begin{figure}[t]
    \centering
    \includegraphics[width=0.45\textwidth]{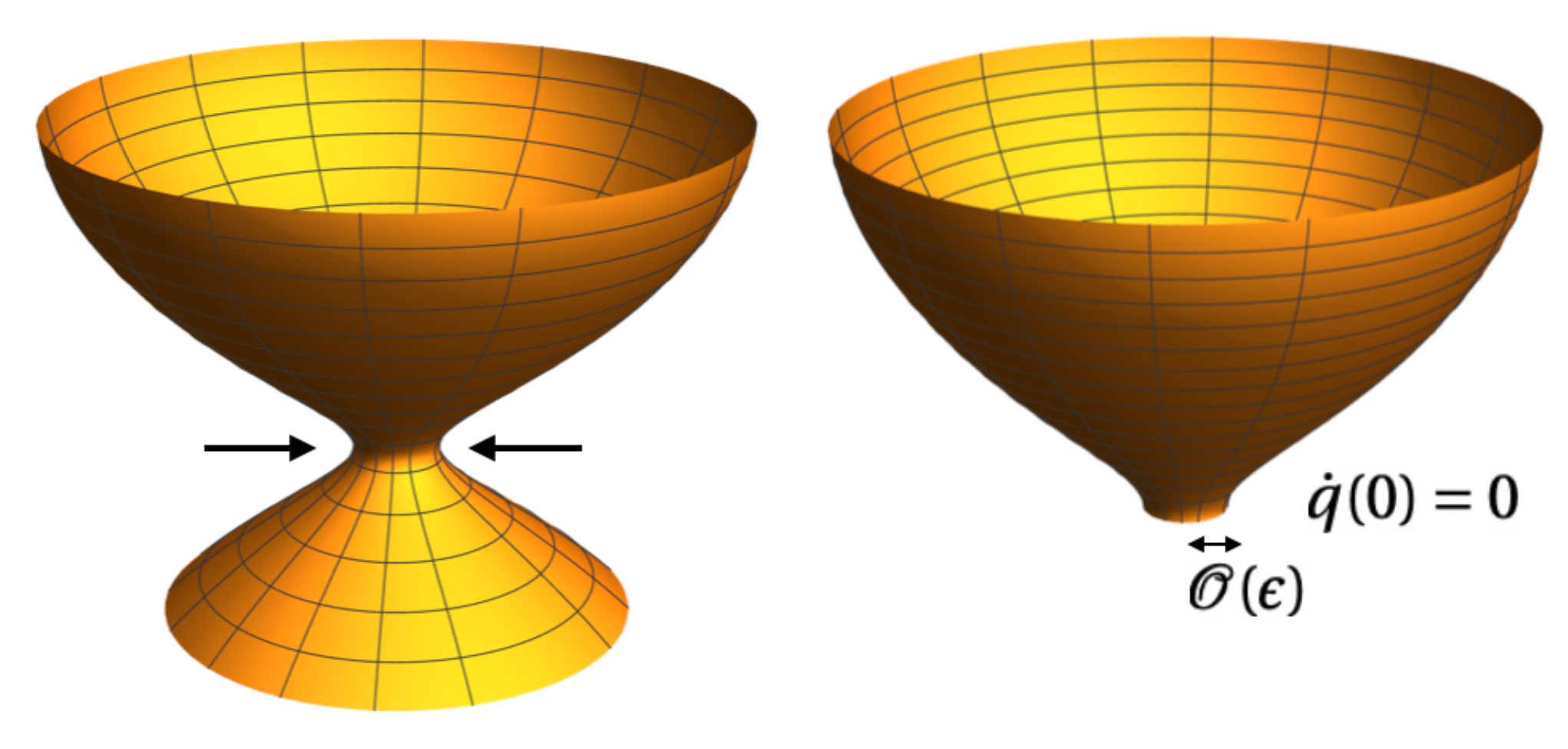}
    \caption{
    Schematic illustration of the small-throat boundary condition.  The left
    panel represents a tunneling geometry connected to another spacetime
    through a small throat.  The parent side may be an asymptotically AdS
    region, another cosmological region, or a more general auxiliary
    spacetime.  In the effective description used in this paper, we keep only
    the universe-side half-geometry and mimic the small throat by imposing a
    boundary condition at the initial endpoint, as shown in the right panel.
    The $\mathcal{O}(\epsilon)$ label refers to the squared scale factor $q=a^2$, rather than to the throat radius $a$.}
    \label{fig:schematic}
\end{figure}

A complementary physical viewpoint is that universe creation may be obtained
as a limiting tunneling process from a very small but finite universe
\cite{Hong:2003pe,Vilenkin:2018dch}.  Such a picture is also suggested by
tunneling or wormhole geometries in which a closed cosmological region is
connected to another spacetime through a throat
\cite{Lavrelashvili:2026zsw,Lavrelashvili:2026jcw} (see Fig.~\ref{fig:schematic} for a schematic picture and see also
Refs.~\cite{Giddings:1987cg,Lavrelashvili:1988un,Rubakov:1988wx,
Jonas:2023ipa,Betzios:2024oli,Lan:2024gnv,Betzios:2024zhf,
Betzios:2026rbv} for related Euclidean wormhole, throat, and baby-universe
constructions.)  If the throat size is controlled by some matter component or
charge, then a decoupling limit may pinch the throat off and leave a closed
universe created from a vanishingly small universe.

In this paper we formulate this idea by imposing a specific boundary condition at the throat and describing the tunneling process as a boundary-value problem, rather than treating the other side of the geometry explicitly. 
For this purpose, we consider a simple closed minisuperspace model
with a positive cosmological constant and a small radiation component.  We use
the variable $q=a^2$, where $a$ is the scale factor, and parametrize the radiation component by a small
positive parameter $\epsilon$.  The radiation term makes $q=0$ singular at
finite $\epsilon$, but it also creates two turning points,
$q_-\simeq \epsilon$ and $q_+\simeq H^{-2}$. 
We interpret the inner turning point $q_-$ as the finite size of the small
throat, while $q_+$ is the usual outer de Sitter turning point.  The desired
tunneling saddle action is obtained only after defining the
finite-$\epsilon$ tunneling problem and taking the limit $\epsilon\to0$.

The small-throat prescription has two ingredients.  First, the initial endpoint
$q_i=q(0)$ is not fixed by a Dirichlet condition, but is integrated over.  As a
result, the saddle geometry satisfies the Neumann condition%
\footnote{
Here the word ``Neumann'' is used in the minisuperspace variational sense. 
At a saddle point of the integral over the initial endpoint, the path obeys
$\dot q(0)=0$.  This should be distinguished from the Neumann and Robin
prescriptions used in Lorentzian no-boundary constructions~\cite{DiTucci:2019dji,DiTucci:2019bui,DiTucci:2020weq}, where one fixes an
initial canonical momentum 
in order to select the desired saddle.  
In a covariant gravitational variational problem, Neumann conditions are more properly understood as boundary-momentum or extrinsic-curvature conditions, rather than as conditions on the coordinate velocity itself~\cite{Krishnan:2016mcj}.  
In the present minisuperspace prescription,
$\dot q(0)=0$ follows from varying the initial endpoint at the saddle,
while the choice of the small-throat contour is additional physical input.
}
\begin{equation}
    \dot q(0)=0 .
\end{equation}
Second, the integration over $q_i$ is restricted to a small relative cycle
$D_\epsilon$ in the punctured disk
\begin{equation}
    0<|q_i|<R_\epsilon,
    \qquad
    R_\epsilon=\sqrt{q_-q_+}=\frac{\sqrt{\epsilon}}{H}.
\end{equation}
This disk contains the inner turning point $q_-$ but excludes the outer turning
point $q_+$.  
Within this punctured disk, we choose the steepest-descent contour through the tunneling saddle. 
This small-domain integration may also be viewed as a sharply supported
initial wave function for the scale factor.  In this respect it is related in
spirit to prescriptions in which an initial wave function is used to regulate
the Lorentzian path integral \cite{Yamada:2025rld}.  In our case, however, the
small support is tied to a finite radiation-induced throat and is supplemented
by the Neumann condition.

Numerically, for the parameters studied below, the
steepest-descent contour associated with the tunneling saddle maps to a contour $D_\epsilon$
in the punctured disk $0 < |q_i| < R_\epsilon$.  This contour starts at $q_i=0$,
passes through $q_i=q_-$, and returns to $q_i=0$ from a different asymptotic
direction.
The zero-radiation limit is taken only after this finite-$\epsilon$ saddle
problem has been defined.  In the limit $\epsilon\to0$, the inner turning point
approaches $q_i=0$, while the tunneling
exponent and the Lorentzian WKB phase reduce to their pure de Sitter values.
Thus the standard tunneling saddle action is recovered as a small-throat
decoupling limit.  
The logic is therefore different from imposing a boundary
condition directly at the degenerate geometry.  The saddle action is first
calculated in a nondegenerate finite-throat problem and only then taken to
the creation-from-nothing limit.%
\footnote{
In a broad sense, 
this use of an auxiliary limiting family may be analogous to
approaches in which the wave function is defined by analytic continuation from
a better-controlled problem, such as from negative to positive potentials
\cite{Lehners:2021jmv}.  In the present paper the auxiliary family is instead
a finite-radiation, finite-throat tunneling problem.
}

The remainder of the paper is organized as follows.  In
Sec.~\ref{sec:small-throat-model}, we introduce the radiation-deformed
minisuperspace model and formulate the small-throat prescription.  In
Sec.~\ref{sec:riemann-sheet}, we derive the on-shell action and specify
the branch of the lapse action used in the prescription.  In
Sec.~\ref{sec:saddles}, we analyze the saddle points, the tunneling exponent,
and the zero-radiation limit.  In Sec.~\ref{sec:numerical}, we present
numerical results for the structure of the complex lapse plane under our
prescription.  We conclude in Sec.~\ref{sec:discussion}.

\section{Radiation minisuperspace and the small-throat prescription}
\label{sec:small-throat-model}

\subsection{Action and turning points}

We consider a closed Friedmann universe with a positive cosmological constant and a homogeneous radiation component.  We use the variable
\begin{equation}
    q(t)=a^2(t)
\end{equation}
and choose the minisuperspace metric~\cite{Louko:1987wq,Halliwell:1988ik}
\begin{equation}
    ds^2
    =
    -\frac{N^2}{q(t)}dt^2
    +
    q(t)d\Omega_3^2,
    \qquad
    0\leq t\leq1 .
    \label{eq:metric}
\end{equation}
Here $N$ is the lapse, which will be complexified in the Picard--Lefschetz analysis.  In units $8\pi G=\hbar=1$, the action is
\begin{equation}
    S_\epsilon[q,N]
    =
    6\pi^2
    \int_0^1dt
    \left[
        -\frac{\dot q^2}{4N}
        +
        N U(q)
    \right],
    \label{eq:action}
\end{equation}
where
\begin{align}
    U(q)
    &=
    1-H^2q-\frac{\epsilon}{q},
    \label{eq:Udef}
    \\
    &=
    H^2
    \frac{(q_+-q)(q-q_-)}{q}.
    \label{eq:Ufactor}
\end{align}
The parameter $\epsilon>0$ represents the radiation component, and $H$ represents the Hubble parameter associated with the vacuum energy.
The turning points $q_\pm$ are given by 
\begin{equation}
    q_\pm
    =
    \frac{
        1\pm\sqrt{1-4H^2\epsilon}
    }{
        2H^2
    },
    \label{eq:qpm}
\end{equation}
for $0<4H^2\epsilon<1$. 
For small $\epsilon$, these give 
\begin{equation}
    q_-\simeq\epsilon,
    \qquad
    q_+\simeq H^{-2}.
\end{equation}
Thus $q_-$ gives a finite but arbitrarily small throat size in the small-$\epsilon$ limit. 

It is useful to introduce the scale
\begin{equation}
    R_\epsilon
    \equiv
    \sqrt{q_-q_+}
    =
    \frac{\sqrt{\epsilon}}{H}.
    \label{eq:Repsilon}
\end{equation}
Since $q_-<q_+$, we have
\begin{equation}
    q_-<R_\epsilon<q_+ .
    \label{eq:qminus-Reps-qplus}
\end{equation}
Thus the disk $|q_i|<R_\epsilon$ contains the inner turning point $q_-$ but excludes the outer turning point $q_+$.

\subsection{Small-throat boundary prescription}
\label{sec:boundary-prescription}

We consider the path integral for the variables $q$ and $N$ with specified boundary conditions. 
We take $t \in [0,1]$ without loss of generality. 
To define the path integral, we fix 
the final size of the universe as
\begin{equation}
    q(1)=q_1,
    \qquad
    q_1>q_+ .
    \label{eq:q1}
\end{equation}
The initial endpoint is not fixed by a Dirichlet condition.  Instead, it is integrated over a small contour $D_\epsilon$ in the complex $q_i$ plane, where
\begin{equation}
    q_i\equiv q(0).
\end{equation}
The small-throat prescription is to choose the initial integration cycle $D_\epsilon$ as a relative one-cycle in the punctured disk
\begin{equation}
    0<|q_i|<R_\epsilon .
\end{equation}
More explicitly, $D_\epsilon$ is a contour satisfying
\begin{equation}
    D_\epsilon:\quad
    q_i(\lambda),
    \qquad
    0<\lambda<1,
    \label{eq:Depsilon-param}
\end{equation}
with
\begin{equation}
    \lim_{\lambda\to0^+}q_i(\lambda)=0,
    \qquad
    \lim_{\lambda\to1^-}q_i(\lambda)=0,
    \label{eq:Depsilon-endpoints}
\end{equation}
while the two limits approach the puncture $q_i=0$ from different angular directions.  
The contour should remain inside the small disk,
\begin{equation}
    |q_i(\lambda)|<R_\epsilon
    \qquad
    \text{for}
    \qquad
    0<\lambda<1 .
    \label{eq:Depsilon-bound}
\end{equation}
Note that this disk includes the inner turning point $q_i = q_-$. 
As we will see below, a relevant saddle corresponds to a contour passing through the inner turning point,
\begin{equation}
    q_i(\lambda_*)=q_-,
    \qquad
    0<\lambda_*<1. 
    \label{eq:Depsilon-qminus}
\end{equation}
At finite $\epsilon$, the point $q_i=0$ is a singular point and is not included as an ordinary point of the contour.  Thus $D_\epsilon$ should be understood as a relative cycle: its two ends approach $q_i=0$ along different branches where the classical exponential factor decays. 
It appears as a closed curve from the origin back to the origin in the $q_i$ plane, but the two ends are distinct on the chosen Riemann sheet.  
For the steepest descent contour, 
they approach $\arg q_i=0$ from below and $\arg q_i=2\pi/3$, respectively, as we will see in Sec.~\ref{sec:branch-continuation}.  The disk alone does not specify this contour. The relative-cycle setting is given in Sec.~\ref{sec:branch-continuation}.

The path integral is then written schematically as
\begin{equation}
    \Psi(q_1)
    =
    \int_{D_\epsilon}dq_i
    \int_{\mathcal C_N}dN
    \int_{q(0)=q_i}^{q(1)=q_1}
    \mathcal Dq\,
    \exp\left[iS_\epsilon[q,N]\right].
    \label{eq:path_integral}
\end{equation}
Before imposing the endpoint saddle condition, $q_i$ and $N$ are independent variables. Here $\mathcal C_N$ denotes the lapse contour in the schematic path integral.

At an endpoint saddle, variation of the action with respect to the initial endpoint gives a Neumann condition.  Here we assume that the initial wave function adds no term to the leading saddle action.  More explicitly,
\begin{equation}
    \delta S_\epsilon
    =
    6\pi^2
    \int_0^1dt
    \left[
        \frac{\ddot q}{2N}
        +
        N U'(q)
    \right]\delta q
    +
    \frac{3\pi^2}{N}\dot q(0)\delta q(0),
    \label{eq:variation}
\end{equation}
where we used $\delta q(1)=0$.  Therefore the classical solutions at an endpoint saddle obey
\begin{equation}
    \dot q(0)=0 .
    \label{eq:neumann}
\end{equation}
The bulk equation of motion is
\begin{equation}
    \ddot q
    =
    -2N^2U'(q)
    =
    2N^2
    \left(
        H^2-\frac{\epsilon}{q^2}
    \right).
    \label{eq:eom}
\end{equation}

Note that the prescription is not a bare Neumann condition alone.  It is the combination of the Neumann condition $\dot q(0)=0$ with the relative small-throat cycle $D_\epsilon\subset\{0<|q_i|<R_\epsilon\}$.  This distinction is crucial.  A bare Neumann condition leaves the initial size unspecified and allows the outer-turning-point saddle $q_i=q_+$.  By contrast, the present prescription includes the inner saddle $q_i=q_-$ and automatically excludes $q_i=q_+$.

For the semiclassical prescription, we choose a steepest-descent lapse contour $\mathcal C_T$ through the inner tunneling saddle, with an outgoing WKB wave and a decaying under-barrier factor, as we will see shortly.  In Sec.~\ref{sec:numerical}, we examine whether it can be deformed toward the positive real lapse axis, but we do not prove that the corresponding steepest-descent contour can be obtained by deforming the Lorentzian path-integral contour. 
In this sense, our prescription is formulated in the semiclassical approximation.

We define the semiclassical prescription used in this paper by
\begin{equation}
 \Psi(q_1)\simeq\int_{\mathcal C_T}dN\,a_\epsilon(N)e^{iS_\epsilon^{\rm cl}(N)},
 \label{eq:min-reduced-prescription}
\end{equation}
where $S_\epsilon^{\rm cl}$ is evaluated on the classical solutions obeying
$\dot q(0)=0$.  The factor $a_\epsilon$ includes the Gaussian fluctuations
and normalization, but the main result below concerns the saddle action.
We have not established the global equivalence of Eq.~\eqref{eq:min-reduced-prescription} to the schematic path integral in Eq.~\eqref{eq:path_integral}, including their integration contours and saddle coefficients.
We take only the steepest-descent contour through the inner tunneling saddle,
with its two ends in regions where $\operatorname{Re}(iS_\epsilon^{\rm cl})$
tends to $-\infty$.  
We have not shown that the finite throat or the parent geometry selects the chosen contour uniquely. We therefore take this contour choice as part of the boundary prescription. The limit calculated below concerns the saddle action.

\section{On-shell action and the selected Riemann sheet}
\label{sec:riemann-sheet}

The path integral over $q$ in 
Eq.~(\ref{eq:path_integral}) can be approximated by using saddle point analysis. 
For fixed $q_i$ and $N$, the classical action can be expressed as a contour integral in the complex $q$ plane. The square roots in the integrand have branch points, so the contour and the square-root branch must be specified.
In this section, we specify which integration contour should be chosen in our prescription when evaluating the classical action.

\subsection{First integral}

The equation of motion admits the first integral
\begin{equation}
    C
    =
    \frac{\dot q^2}{4N^2}
    +
    U(q).
    \label{eq:first_integral}
\end{equation}
Therefore
\begin{equation}
    \dot q
    =
    2N
    \sqrt{
        C-U(q)
    },
    \label{eq:qdot}
\end{equation}
where the square-root branch is part of the saddle specification.

The condition that the solution reaches
$q_1$ at $t=1$ is
\begin{equation}
    2N
    =
    \int_{\mathcal C_q(q_i,q_1)}
    \frac{dq}{
        \sqrt{C-U(q)}
    },
    \label{eq:N_quadrature}
\end{equation}
where $\mathcal C_q(q_i,q_1)$ is an integration contour from $q=q_i$ to
$q=q_1$ in the complex $q$ plane.  For a chosen branch of the classical
solution, this equation determines $C$ for the given $q_i$ and $N$.

Using the Neumann condition $\dot q(0)=0$ and writing
\begin{equation}
    q_i=q(0),
\end{equation}
we obtain
\begin{equation}
    C
    =
    U(q_i)
    =
    1-H^2q_i-\frac{\epsilon}{q_i}.
    \label{eq:C_Uqi}
\end{equation}
Equation~\eqref{eq:N_quadrature} then applies with 
$C=U(q_i)$.  Thus the lapse for these classical solutions is naturally
parametrized by the initial value $q_i$.

\subsection{On-shell action}

At fixed $q_i$ and $N$, we denote the classical action by $S_D(q_i,N;q_1)$.
The on-shell action is obtained by using the first integral.  Since
\begin{equation}
    -\frac{\dot q^2}{4N}+NU(q)
    =
    N\left(2U(q)-C\right),
\end{equation}
and
\begin{equation}
    dt
    =
    \frac{dq}{
        2N\sqrt{C-U(q)}
    },
\end{equation}
we find
\begin{align}
    S_D(q_i,N;q_1)
    &=
    3\pi^2
    \int_{\mathcal C_q(q_i,q_1)}
    dq\,
    \frac{
        2U(q)-C
    }{
        \sqrt{C-U(q)}
    }
    \label{eq:Scl1}
\\
    &=
    6\pi^2 C N
    -
    6\pi^2
    \int_{\mathcal C_q(q_i,q_1)}
    dq\,
    \sqrt{C-U(q)},
    \label{eq:Scl2}
\end{align}
where we used \eq{eq:N_quadrature} in the second line.
Varying the initial endpoint and the lapse independently gives
\begin{align}
 \frac{\partial S_D}{\partial q_i}
 &=\frac{3\pi^2}{N}\dot q(0),\nonumber\\
 \frac{\partial S_D}{\partial N}&=6\pi^2C.
 \label{eq:min-dirichlet-HJ}
\end{align}
As in Sec.~\ref{sec:boundary-prescription}, the endpoint saddle condition
$\partial S_D/\partial q_i=0$ therefore gives $\dot q(0)=0$.
Only at this stage do we write $q_i=q_i(N)$ and obtain $C=U(q_i)$.%
\footnote{\label{fn:local-gaussian}
Near an isolated endpoint saddle, the Gaussian approximation can also be
checked for this order of calculation.  Writing subscripts for partial
derivatives and assuming $(S_D)_{q_i q_i}\ne0$, we find
\begin{align*}
 \frac{d^2S_\epsilon^{\rm cl}}{dN^2}
 &=(S_D)_{NN}-\frac{(S_D)_{N q_i}(S_D)_{q_i N}}{(S_D)_{q_i q_i}},\\
 \det S_D''&=(S_D)_{q_i q_i}\frac{d^2S_\epsilon^{\rm cl}}{dN^2}.
\end{align*}
Here $S_D''$ is the matrix of second derivatives with respect to $q_i$ and
$N$.  Thus integrating first over $q_i$ and then over $N$ gives the same
Gaussian factor as integrating over both variables together near a
nondegenerate saddle with respect to both $q_i$ and $N$, with phases fixed
by the integration contour.  
However, we note that this local check does not determine which
endpoint saddles contribute to the full integral, how many times their
contours occur, or whether the endpoints give additional contributions.
}
The holomorphic exponent for the lapse integral in
Eq.~\eqref{eq:min-reduced-prescription} is then
\begin{equation}
    \mathcal I(N)
    =
    iS_{\epsilon}^{\rm cl}(N) =i S_D(q_i(N),N;q_1).
    \label{eq:Idef}
\end{equation}

\subsection{Branch points and the small-throat sheet}
\label{sec:branch-continuation}

For fixed $q_i$, the branch points of the square root in Eq.~\eqref{eq:N_quadrature} are the solutions of
\begin{equation}
    C-U(q)=U(q_i)-U(q)=0 .
\end{equation}
One solution is $q=q_i$.  The other solution, denoted by $q_i'$, is given by 
\begin{equation}
    q_i'
    =
    \frac{\epsilon}{H^2q_i}.
    \label{eq:qiprime}
\end{equation}

\begin{figure}[t]
    \centering
    \includegraphics[width=0.45\textwidth]{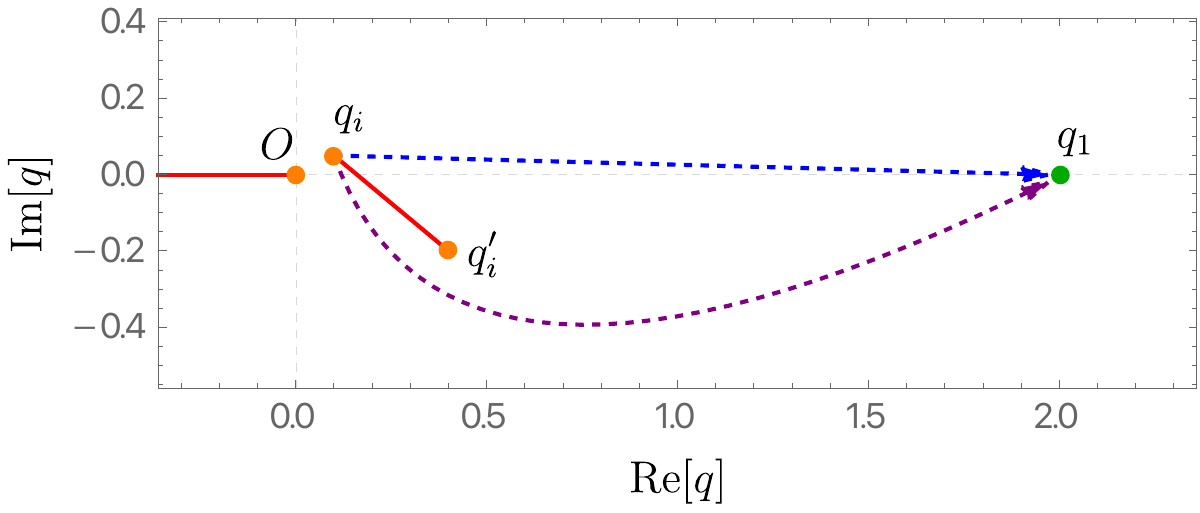}
    \caption{
    Two representative integration contours in the complex $q$ plane.  The red lines denote branch cuts ending at $q=0$, $q=q_i$, and $q=q_i'$.  The blue and purple dashed curves correspond to two different choices of contour around the cut connecting $q_i$ and $q_i'$.  The figure uses $q_1=2$, $\epsilon=0.05$, and $q_i=0.1+0.05i$ as an illustrative example.}
    \label{fig:branch}
\end{figure}

The square root also has a branch point at $q=0$, inherited from the radiation term.  We choose a branch cut to run from $q=0$ to $-\infty$ and another branch cut to connect $q_i$ and $q_i'$.  For ${\rm Re}\,q_i>0$, the latter cut may be taken directly between $q_i$ and $q_i'$ (see Fig.~\ref{fig:branch}).  The contours described below are continued from the tunneling saddle as $q_i$ varies; the sign of $\Re q_i$ alone does not fix the square-root branch.  The explicit formulas below specify this continuation without additional winding around the branch points.
In our numerical implementation, 
we write the contour integral for $N$ in \eq{eq:N_quadrature} as
\begin{equation}
    2N_{s,\eta}(q_i)
    =
    \eta i
    \int_{q_i}^{q_i'}
    \frac{dq}{
        \sqrt{U(q)-U(q_i)}
    }
    +
    s
    \int_{q_i'}^{q_1}
    \frac{dq}{
        \sqrt{U(q_i)-U(q)}
    },
    \label{eq:N_branch}
\end{equation}
for $\Re[q_i] >0$, 
where $s=\pm1$ labels the Lorentzian WKB branch and $\eta=\pm1$ labels the decaying or growing Euclidean branch. 
The blue and purple dashed curves in the figure correspond to the integrations for the cases $(s, \eta) = (1,1)$ and $(s,\eta) = (1,-1)$, respectively. 

By contrast, 
for $\Re[q_i] \le 0$, 
the branch cut intersects the negative real axis, and it should be deformed so that it detours to the positive-real side around the origin.  
In this case, 
the contour integral for $N$ in \eq{eq:N_quadrature} is calculated as
\begin{align}
    2N_{s,\eta}(q_i)
    &=
    \eta i \lmk 
    \int_{q_i}^{0}
    \frac{dq}{
        \sqrt{U(q)-U(q_i)}
    }
    +
    \int_{0}^{q_i'}
    \frac{dq}{
        \sqrt{U(q)-U(q_i)}
    } 
    \rmk
    \nonumber\\
    &+
    s
    \int_{q_i'}^{q_1}
    \frac{dq}{
        \sqrt{U(q_i)-U(q)}
    },
    \label{eq:N_branch2}
\end{align}
for $\Re[q_i] \le 0$.  The point $q=0$ in this expression is not part of the
integration contour; it denotes the limiting endpoint of a small detour around
the puncture on the positive-real side.

Different choices of $\mathcal C_q(q_i,q_1)$ (or of $\eta$ and $s$) give different values of the lapse $N$ and of the corresponding on-shell action.  This is the usual multi-valuedness of the minisuperspace action: different branches correspond to incoming or outgoing WKB waves in the Lorentzian region and to growing or decaying wave functions in the Euclidean region.
In the numerical calculation below, we include all combinations of $\eta = \pm 1$ and $s = \pm 1$ to display these saddle branches in the complex $N$ plane. 
Here we consider these four branches without additional winding around the branch points. Additional winding can give further values of the integrals. The tunneling saddle corresponds to $s=\eta=1$, and we follow this solution continuously as described below. 
When we follow the same tunneling solution as $q_i$ varies, we instead use an alternative method explained in App.~\ref{app:B}.

For $\Re[q_i] >0$, the corresponding on-shell action \eq{eq:Scl2} is evaluated as
\begin{align}
    S_{\epsilon,s,\eta}^{\rm cl}(q_i)
    &=
    6\pi^2U(q_i)N_{s,\eta}(q_i)
    \nonumber\\
    &\quad
    +6\pi^2\eta i
    \int_{q_i}^{q_i'}
    dq\,
    \sqrt{U(q)-U(q_i)}
    \nonumber\\
    &\quad
    -6\pi^2s
    \int_{q_i'}^{q_1}
    dq\,
    \sqrt{U(q_i)-U(q)} .
    \label{eq:Scl_branch}
\end{align}
The case for $\Re[q_i] \le 0$ is evaluated similarly to \eq{eq:N_branch2}.

Note that, 
using $q_-q_+=\epsilon/H^2$, \eq{eq:qiprime} can be written as
\begin{equation}
    q_i'
    =
    \frac{q_-q_+}{q_i}
    =
    \frac{R_\epsilon^2}{q_i}.
    \label{eq:qiprime-Reps}
\end{equation}
Thus the map $q_i\mapsto q_i'$ is an inversion with respect to the circle $|q_i|=R_\epsilon$.  If $q_i$ lies inside the disk $|q_i|<R_\epsilon$, then its partner branch point $q_i'$ lies outside this disk.  In particular, the inner turning point and the outer turning point are exchanged by this inversion:
\begin{equation}
    q_i=q_-
    \qquad
    \Longrightarrow
    \qquad
    q_i'=q_+ .
\end{equation}
This observation explains why the disk $|q_i|<R_\epsilon$ is the natural small-throat region.  It contains the tunneling saddle $q_i=q_-$, while the unsuppressed outer saddle $q_i=q_+$ lies outside it.  Moreover, for $q_i$ in this disk, the branch cut connecting $q_i$ and $q_i'$ connects an inner branch point to an outer branch point.  This is the branch structure used for the tunneling geometry connecting the small-throat side to the expanding-universe side.

The point $q_i=0$ is special.  At finite $\epsilon$,
\begin{equation}
    U(q_i)
    =
    1-H^2q_i-\frac{\epsilon}{q_i}
\end{equation}
has a pole at $q_i=0$, and Eq.~\eqref{eq:qiprime-Reps} sends the second branch point to infinity:
\begin{equation}
    q_i\to0
    \qquad
    \Longrightarrow
    \qquad
    q_i'\to\infty .
\end{equation}
Thus $q_i=0$ is a puncture of the family of classical solutions, not an ordinary point of the $q_i$ plane.  The two ends of the small-throat contour $D_\epsilon$ approach this point on different branches.

Near the puncture, the classical action is not single-valued as an ordinary function of $q_i$.  At fixed $\epsilon>0$, the two ends of the selected continuation have different asymptotics.  We call the branch with $N\to0$ the short branch, and the one with $|N|\to\infty$ the long branch.  On the short branch,
\begin{align}
    N(q_i)&\sim\frac{iq_1\sqrt{q_i}}{2\sqrt\epsilon},\nonumber\\
    iS_{\epsilon}^{\rm cl}(q_i)&\sim-\frac{3\pi^2q_1\sqrt\epsilon}{\sqrt{q_i}},
    \label{eq:qi-origin-short}
\end{align}
whereas on the long branch,
\begin{align}
    N(q_i)&\sim\frac{2i\sqrt\epsilon}{H^2\sqrt{q_i}},\nonumber\\
    iS_{\epsilon}^{\rm cl}(q_i)&\sim\frac{4\pi^2\epsilon^{3/2}}{H^2q_i^{3/2}}.
    \label{eq:qi-origin-asymptotic}
\end{align}
The long-branch powers follow by rescaling $q=q_i'x$.  On the short branch the leading long-segment terms cancel, so the same powers cannot be assigned to both ends.
For $q_i=re^{i\theta}$ with $\theta$ followed continuously along the contour, exponential decay requires $\cos(\theta/2)>0$ on the short branch and $\cos(3\theta/2)<0$ on the long branch.  
As we will see shortly, 
the selected ends for the steepest descent contour approach $\theta=0$ from below and $\theta=2\pi/3$, respectively, which map to $N\to+i0$ and $N\to\infty$.
Thus a contour can appear in the projected $q_i$ plane as a closed curve from the origin to the origin, while its two ends lie on different branches.

Subleading terms include $q_i^{3/2}\log q_i$ in the lapse.  Thus winding around $q_i=0$ can change more than the sign of the leading square root.  The contour integrals can also change by integrals around the branch cuts. 
The above prescription and Eq.~\eqref{eq:min-branch-atlas} specify the continuation used here, without additional winding.

More explicitly, we denote by $\Sigma_\epsilon$ the Riemann
surface of the pair $(N,S_\epsilon^{\rm cl})$ obtained by
analytic continuation from the inner tunneling saddle,
restricted to $0<|q_i|<R_\epsilon$ and excluding singular points.
The short and long ends lie in the decaying sectors specified above, on this surface rather than just in the projected $q_i$ plane.
For large $M>0$, we define $\mathcal E_{\epsilon,M}$ as the
union of small neighborhoods of the short and long ends
within the decaying sectors specified above, restricted to
$\operatorname{Re}(iS_\epsilon^{\rm cl})<-M$.
After truncating its two ends in $\mathcal E_{\epsilon,M}$, the lifted contour $D_\epsilon$ represents a class in the relative homology group
\begin{align}
 H_1(\Sigma_\epsilon,\mathcal E_{\epsilon,M};\mathbb Z).
\end{align}
Here relative means that the ends may move within $\mathcal E_{\epsilon,M}$ during a continuous deformation.
The limit $M\to\infty$ describes the decaying ends at the puncture.
We choose the class represented by the lifted image of $\mathcal C_T$, using the continuation in App.~\ref{app:B} without adding further winding around the puncture or the branch points of the $q$ integrals.
We do not classify all classes or claim that this choice is unique.

\section{Saddle points and tunneling wave function}
\label{sec:saddles}

We now analyze the saddle points for the $N$ integral in Eq.~\eqref{eq:min-reduced-prescription} and show that the small-throat saddle action reduces to that of the standard tunneling wave function in the limit $\epsilon\to0$.

\subsection{Lapse saddle condition}

The lapse saddle condition follows from the on-shell action.  Since the initial Neumann condition and the fixed final value make the boundary terms vanish,
\begin{equation}
    \frac{dS_{\epsilon}^{\rm cl}}{dN}
    =
    \left.
    \frac{\partial S_\epsilon}{\partial N}
    \right|_{q=q_{\rm cl}}.
\end{equation}
Using Eq.~\eqref{eq:first_integral}, we obtain
\begin{equation}
    \frac{dS_{\epsilon}^{\rm cl}}{dN}
    =
    6\pi^2
    \int_0^1dt
    \left[
        \frac{\dot q^2}{4N^2}
        +
        U(q)
    \right]
    =
    6\pi^2C.
    \label{eq:dSdN}
\end{equation}
Therefore the lapse saddle condition is
\begin{equation}
    C=0.
    \label{eq:Czero}
\end{equation}
Since $C=U(q_i)$, the saddle values of $q_i$ are the turning points:
\begin{equation}
    U(q_i)=0
    \qquad
    \Longrightarrow
    \qquad
    q_i=q_\pm.
    \label{eq:qi_saddles}
\end{equation}

A bare Neumann boundary condition would allow both roots $q_i=q_-$ and $q_i=q_+$.  In the small-throat prescription, however, the chosen lapse contour maps into the punctured disk
\begin{equation}
    0<|q_i|<R_\epsilon,
    \qquad
    R_\epsilon=\sqrt{q_-q_+}.
\end{equation}
Since
\begin{equation}
    q_-<R_\epsilon<q_+,
\end{equation}
the inner saddle $q_i=q_-$ lies in the small-throat domain, while the outer saddle $q_i=q_+$ is outside it.  Together with this domain, we specify the integration contour through the inner tunneling saddle,
\begin{equation}
    q_i=q_-, 
\end{equation}
and correspondingly $q_i' = q_+$.

\subsection{Small-throat tunneling saddle}

Here we calculate $N$ and $S$ at the $N$ saddle point in terms of the elliptic integrals. 

Define the Euclidean and Lorentzian time integrals
\begin{equation}
    T_E
    =
    \int_{q_-}^{q_+}
    \frac{dq}{\sqrt{U(q)}},
    \qquad
    T_L(q_1)
    =
    \int_{q_+}^{q_1}
    \frac{dq}{\sqrt{-U(q)}}.
    \label{eq:TE_TL}
\end{equation}
Then the four lapse saddles without additional winding are given by 
\begin{equation}
    N_{-,*}^{(s,\eta)}
    =
    \frac{s}{2}T_L(q_1)
    +
    \frac{i\eta}{2}T_E,
    \qquad
    s=\pm1,
    \qquad
    \eta=\pm1.
    \label{eq:Nstar_minus}
\end{equation}
We also define the under-barrier action and the Lorentzian phase by
\begin{equation}
    B_\epsilon
    =
    6\pi^2
    \int_{q_-}^{q_+}
    dq\,
    \sqrt{U(q)},
    \label{eq:B_epsilon}
\end{equation}
and
\begin{equation}
    \Phi_\epsilon(q_1)
    =
    6\pi^2
    \int_{q_+}^{q_1}
    dq\,
    \sqrt{-U(q)}.
    \label{eq:Phi_epsilon}
\end{equation}
For the small-throat saddles one finds
\begin{equation}
    iS_{-,*}^{(s,\eta)}
    =
    -\eta B_\epsilon
    -
    is\,\Phi_\epsilon(q_1), 
    \label{eq:iS_small}
\end{equation}
from \eq{eq:Scl_branch}. 
In the branch convention used below, the tunneling contribution is obtained from $\eta=s=1$, for which
\begin{equation}
    \exp\left[iS_{-,*}^{(+,+)}\right]
    =
    \exp[-B_\epsilon]\,
    \exp[-i\,\Phi_\epsilon(q_1)].
    \label{eq:tunneling_contribution}
\end{equation}
The branch $\eta=-1$ gives $\exp[+B_\epsilon]$ and corresponds to the growing under-barrier weighting characteristic of Hartle--Hawking-type contours. 

The tunneling exponent can be expressed in closed form.  Using Eq.~\eqref{eq:Ufactor}, we can write 
\begin{equation}
    B_\epsilon
    =
    6\pi^2H
    \int_{q_-}^{q_+}
    dq\,
    \sqrt{
        \frac{(q_+-q)(q-q_-)}{q}
    }.
    \label{eq:B_factor}
\end{equation}
We define 
\begin{equation}
    m_\epsilon
    =
    1-\frac{q_-}{q_+}
    =
    \frac{q_+-q_-}{q_+}, 
    \label{eq:m_epsilon}
\end{equation}
and obtain 
\begin{equation}
    T_E
    =
    \frac{2\sqrt{q_+}}{H}
    E(m_\epsilon),
    \label{eq:TE_elliptic}
\end{equation}
and
\begin{equation}
    B_\epsilon
    =
    4\pi^2H\sqrt{q_+}
    \left[
        (q_++q_-)E(m_\epsilon)
        -
        2q_-K(m_\epsilon)
    \right], 
    \label{eq:B_elliptic}
\end{equation}
where $K(m)$ and $E(m)$ denote complete elliptic integrals of the first and second kind, respectively.

\subsection{Outer-turning-point saddle}

It is useful to contrast the above result with the saddle structure of a bare Neumann problem.  If one imposes only $\dot q(0)=0$, then the saddle condition $C=0$ allows both $q_i=q_-$ and $q_i=q_+$.  The second choice corresponds to an outer-turning-point saddle.  For $q_i=q_+$, no under-barrier segment exists, and the lapse saddle is
\begin{equation}
    N_{+,*}^{(s)}
    =
    \frac{s}{2}T_L(q_1),
    \qquad
    s=\pm1 .
    \label{eq:Nstar_plus}
\end{equation}
The exponent is given by 
\begin{equation}
    iS_{+,*}^{(s)}
    =
    -is\,\Phi_\epsilon(q_1).
    \label{eq:iS_plus}
\end{equation}
Thus this saddle is not suppressed by the factor $\exp[-B_\epsilon]$.

This is precisely why a bare Neumann boundary condition is not sufficient to define a tunneling-from-small-universe wave function.  In the present prescription, however, $D_\epsilon$ is the image of the chosen lapse contour through the inner tunneling saddle.  The outer saddle $q_i=q_+$ is excluded by this choice.

\subsection{Zero-radiation limit}

At fixed $q_1>H^{-2}$, away from the final turning point, the limit $\epsilon\to0$ is taken only after the finite-$\epsilon$ saddle problem has been defined.  In this limit,
\begin{equation}
    q_-\to0,
    \qquad
    q_+\to H^{-2},
    \qquad
    R_\epsilon\to0.
    \label{eq:epsilon_limit}
\end{equation}
The tunneling exponent becomes
\begin{equation}
    B_\epsilon
    \to
    6\pi^2
    \int_0^{H^{-2}}
    dq\,
    \sqrt{1-H^2q}
    =
    \frac{4\pi^2}{H^2}.
    \label{eq:B_limit}
\end{equation}
The Lorentzian phase becomes
\begin{equation}
    \Phi_\epsilon(q_1)
    \to
    6\pi^2
    \int_{H^{-2}}^{q_1}
    dq\,
    \sqrt{H^2q-1}
    =
    \frac{4\pi^2}{H^2}
    \left(
        H^2q_1-1
    \right)^{3/2}.
    \label{eq:Phi_limit}
\end{equation}
Therefore the leading exponential factor of the small-throat tunneling saddle gives
\begin{equation}
    \exp\bigl[iS_{-,*}^{(+,+)}\bigr]
    \longrightarrow
    \exp\left[
        -\frac{4\pi^2}{H^2}
        -
        i
        \frac{4\pi^2}{H^2}
        \left(
            H^2q_1-1
        \right)^{3/2}
    \right],
    \label{eq:tunneling_limit}
\end{equation}
This is the exponent and outgoing WKB phase of the standard tunneling wave function in the pure de Sitter limit~\cite{Feldbrugge:2017kzv,Feldbrugge:2017fcc}.

The limiting behavior in \eq{eq:epsilon_limit} is simple for the saddle action.
A contour that stays inside $|q_i|<R_\epsilon$ would also shrink to the
origin as $\epsilon\to0$.  
The saddle action therefore gives the tunneling exponent and outgoing WKB phase in the same limit. 
Thus, although we start from a Neumann-like boundary condition, we obtain the leading exponential factor of the wave function of the universe from nothing by imposing the supplemental condition
$q_i\in D_\epsilon\subset\{0<|q_i|<R_\epsilon\}$. 
This is the small-throat boundary prescription. 

However, we should note that Equation~\eqref{eq:tunneling_limit}
is obtained directly from the saddle action. To establish the limit of the
full wave function, one must also calculate the prefactor and normalization
(see Appendix~\ref{app:min-measure}) and check the semiclassical approximation
near $q_-$.

\section{Numerical results}
\label{sec:numerical}

\subsection{Steepest-descent contours}
\label{sec:numerical-thimbles}

\begin{figure}[t]
    \centering
    \includegraphics[width=0.43\textwidth]{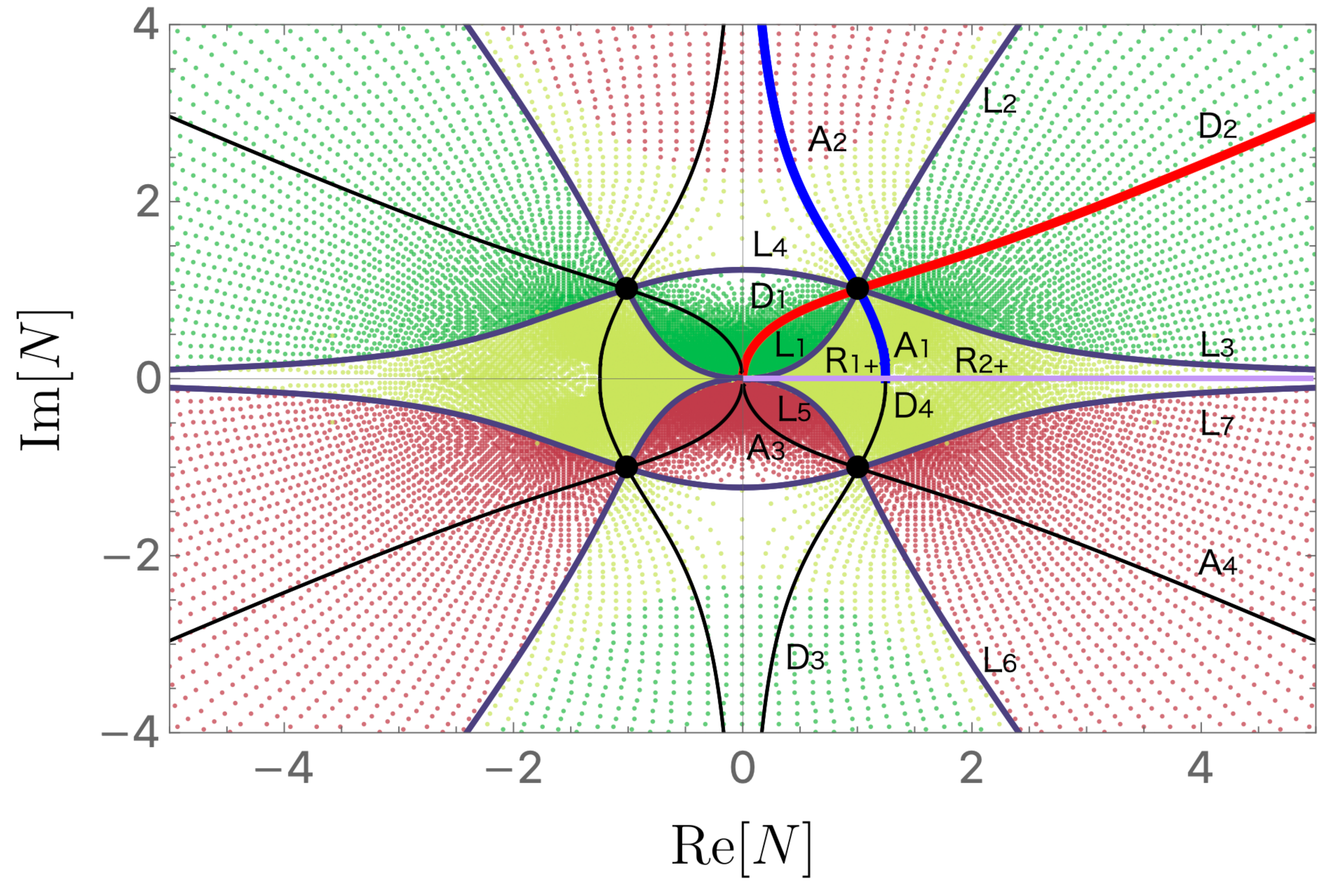}
    \vspace{0.4cm}
    \\
    \includegraphics[width=0.43\textwidth]{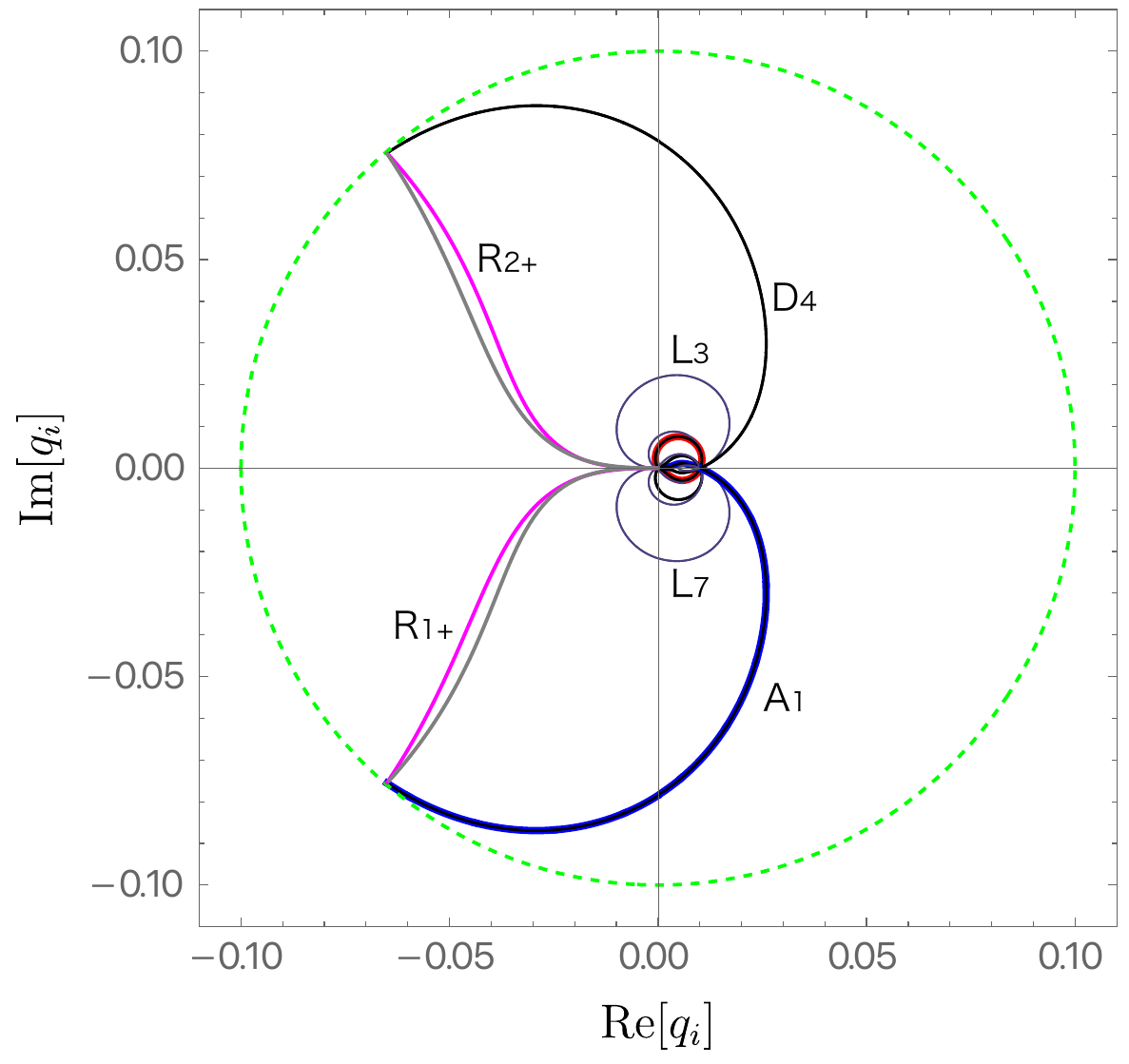}
    \vspace{0.3cm}
    \\
    \includegraphics[width=0.43\textwidth]{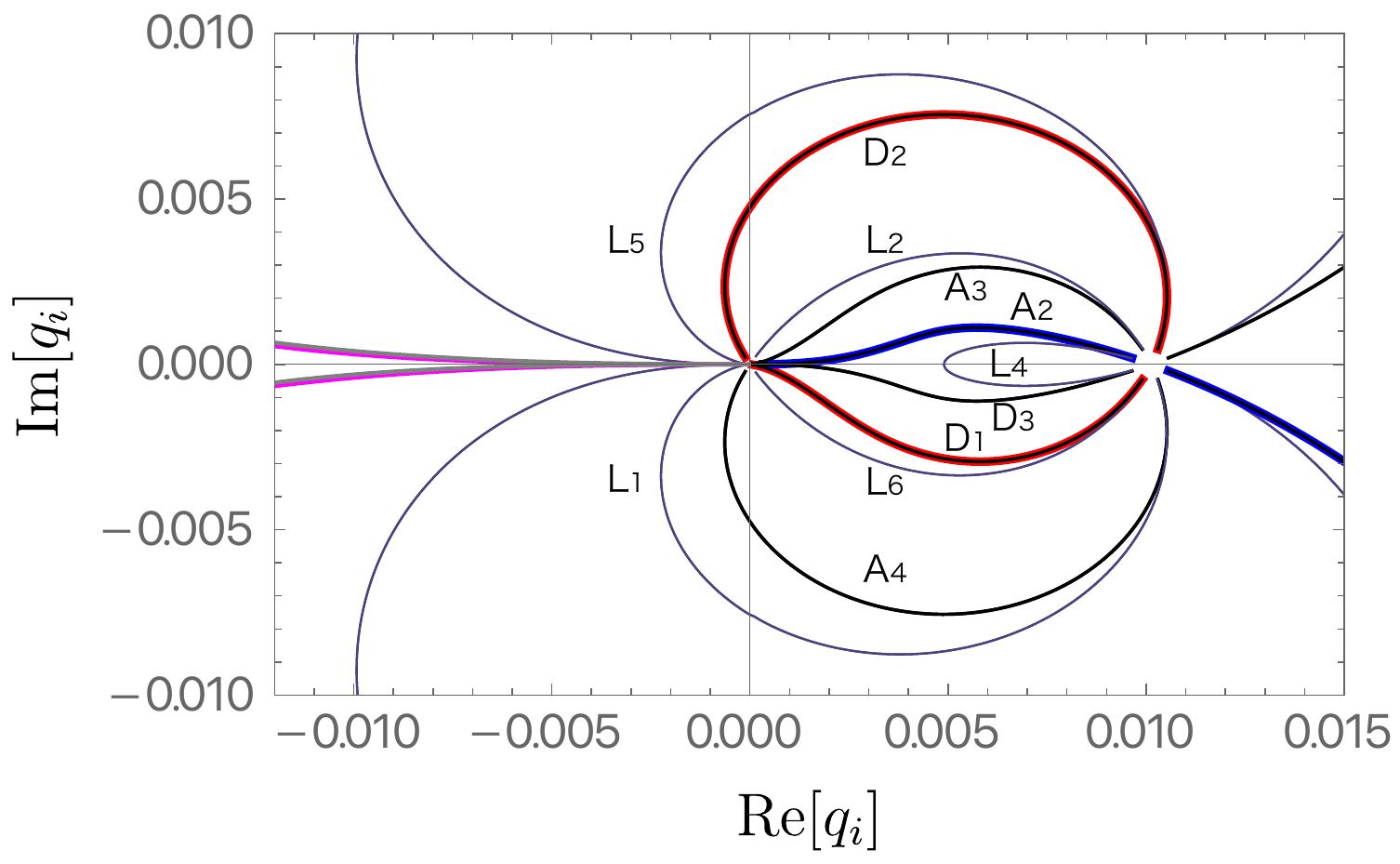}
    \caption{
    Complex lapse plane (upper panel) and corresponding $q_i$ plane (middle and lower panels) for the small-throat prescription.  The plots are evaluated on classical solutions obeying the Neumann condition, with the displayed branch choices.  
    }
    \label{fig:small-throat-plane}
\end{figure}

We now illustrate the small-throat prescription numerically for the classical solutions obeying the Neumann condition.  We evaluate Eqs.~\eqref{eq:N_branch} and \eqref{eq:Scl_branch} for complex values of $q_i$ in the punctured disk
\begin{equation}
    0<|q_i|<R_\epsilon,
    \qquad
    R_\epsilon=\sqrt{q_-q_+} = \frac{\sqrt{\epsilon}}{H}.
    \label{eq:numerical-disk}
\end{equation}
The numerical results illustrate the image of this small disk in the complex lapse plane, within the plotted range, through the multi-valued map
\begin{equation}
    q_i
    \mapsto
    N_{s,\eta}(q_i).
    \label{eq:map}
\end{equation}
In particular, the Picard--Lefschetz contours associated with the tunneling saddle project to relative curves in the $q_i$ plane.  Their projections start at $q_i=0$, pass through the inner turning point $q_i=q_-$, and return to $q_i=0$ from a different angular direction.  Since $q_i=0$ is a puncture at finite $\epsilon$, the two ends are distinct on the chosen Riemann sheet, even though they meet at the origin in the plot.

A useful diagnostic is the relative Morse function
\begin{equation}
    \Delta R(N)
    =
    \operatorname{Re}
    \left[
        iS_{\epsilon}^{\rm cl}(N)
    \right]
    -
    \operatorname{Re}
    \left[
        iS_{\epsilon}^{\rm cl}(N_*)
    \right],
    \label{eq:DeltaR}
\end{equation}
where $N_*$ denotes the tunneling saddle.  Curves satisfying
\begin{equation}
    \operatorname{Im}
    \left[
        iS_{\epsilon}^{\rm cl}(N)
    \right]
    =
    \operatorname{Im}
    \left[
        iS_{\epsilon}^{\rm cl}(N_*)
    \right]
    \label{eq:constant-phase}
\end{equation}
identify the steepest-descent and steepest-ascent contours through the saddle.  The descent branches are those along which $\operatorname{Re}[iS_{\epsilon}^{\rm cl}]$ decreases away from the saddle.

We can set $H = 1$ without loss of generality. 
In Fig.~\ref{fig:small-throat-plane}, we also take
\begin{equation}
    \epsilon=0.01,
    \qquad
    q_1=2,
\end{equation}
as an example. 
Then
\begin{equation}
    R_\epsilon=\sqrt{\epsilon}=0.1,
\end{equation}
while
\begin{equation}
    q_-\simeq0.0101,
    \qquad
    q_+\simeq0.9899.
\end{equation}

The upper panel shows the image of the sampled domain in the complex $N$ plane.
We first sample
\begin{equation}
q_i = r e^{i\theta}
\end{equation}
on a $200\times 100$ grid in the punctured disk $0<|q_i|<R_\epsilon$ for all combinations of $\eta=\pm1$ and $s=\pm1$.
More explicitly, $\log r$ and $\theta$ are uniformly spaced over
\begin{equation}
r\in (10^{-5},R_\epsilon),\qquad \theta\in(0,2\pi),
\end{equation}
Moreover, as a plotting convention we keep the samples satisfying $\Im [N] \ge 0$ for $\eta = 1$
and $\Im [N] \le 0$ for $\eta = -1$.  
This restriction is imposed only on the points displayed in the figure. The Riemann sheet used to continue the action is specified in Sec.~\ref{sec:branch-continuation}.

The color of each point indicates the value of
$\operatorname{Re}[iS_{\epsilon}^{\rm cl}(N)]$ relative to the saddle values.
Green points satisfy
\begin{equation}
\operatorname{Re}[iS_{\epsilon}^{\rm cl}(N)]
<
\operatorname{Re}[iS_{\epsilon}^{\rm cl}(N_{-,*}^{(1,1)})],
\end{equation}
whereas red points satisfy
\begin{equation}
\operatorname{Re}[iS_{\epsilon}^{\rm cl}(N)]
>
\operatorname{Re}[iS_{\epsilon}^{\rm cl}(N_{-,*}^{(1,-1)})].
\end{equation}
Yellow points lie between these two saddle values.  The red curves, labeled
$D_1$ and $D_2$, and the blue curves, labeled $A_1$ and $A_2$, represent the
steepest-descent and steepest-ascent contours associated with the tunneling
saddle $N_{-,*}^{(1,1)}$.  The black curves, including $A_3$, $A_4$, $D_3$,
and $D_4$, show the corresponding constant-phase contours for the other
branches.  
The gray curves, labeled $L_i$ with $i=1,\ldots,7$, are contours of constant $\operatorname{Re}[iS_{\epsilon}^{\rm cl}(N)]$, each passing through one of the saddle points $N_{-,*}^{(\pm1,\pm1)}$.

The middle and lower panels show the corresponding contours in the complex
$q_i$ plane.  We restrict $q_i$ to the punctured disk
\begin{equation}
0<|q_i|<R_\epsilon=\sqrt{\epsilon}/H,
\end{equation}
whose boundary is shown as the dashed green circle in the middle panel.  The
steepest-descent contours attached to the tunneling saddle project to relative
curves in the punctured $q_i$ plane: their projections start near
$q_i=0$, pass through $q_i=q_-$, and return to $q_i=0$ from a different angular
direction.  
The steepest-descent contour extending from the saddle in this example remains inside $|q_i|<R_\epsilon$.
A contour staying in this disk would shrink to the origin as $\epsilon\to0$.

We also identify the preimage of the positive real axis in the complex $N$
plane.  For the branch $s=\eta=1$, the positive real axis, denoted by
$R_{1+}$ and $R_{2+}$, maps to the magenta curves in the $q_i$ plane, running
from the origin to the boundary $|q_i|=R_\epsilon$ in the plotted representations.  On this boundary, the two
branch points are related by complex conjugation, $q_i'\!=q_i^*$.

The gray curves close to the magenta ones in the lower panels represent the
same positive real axis in the complex $N$ plane, but for the branch
$(s,\eta)=(1,-1)$. 
As noted above, 
in the plot we discard the points with 
$\Im [N] < 0$ for $\eta = 1$ 
and $\Im [N] > 0$ for $\eta = -1$.  This removes
the region on the left side of the magenta and gray curves for each choice of $s$ and $\eta$.

\subsection{Deformation toward the real lapse axis}
\label{sec:real-lapse-deformation}

We now start from the chosen steepest-descent contour $D_1+D_2$ and examine its deformation toward the positive real lapse axis, following the corresponding $q_i$ solution continuously.  On the continuation used in App.~\ref{app:B}, we find a point inside the disk where
\begin{align}
N'(q_{i,b})=0,
\end{align}
at 
\begin{align}
 q_{i,b}&\simeq0.047+0.031i,\nonumber\\
 N_b&\equiv N(q_{i,b}) \simeq 1.47+0.61 i.
 \label{eq:interior-Nprime-zero}
\end{align}
Thus the inverse map from $N$ to $q_i$ has a branch point at $N_b$, at which two values of $q_i$ approach the same value. Following $N$ once around $N_b$ changes one solution into the other, and they connect the short and long branches.  To connect these solutions continuously while avoiding $N_b$, this deformation retains a detour.  This point is not a new lapse saddle because $U(q_{i,b}) \ne0$, or $dS_\epsilon^{\rm cl}/dN=6\pi^2U(q_{i,b})\ne0$.

\begin{figure*}[t]
 \centering
 \includegraphics[width=0.91\textwidth]{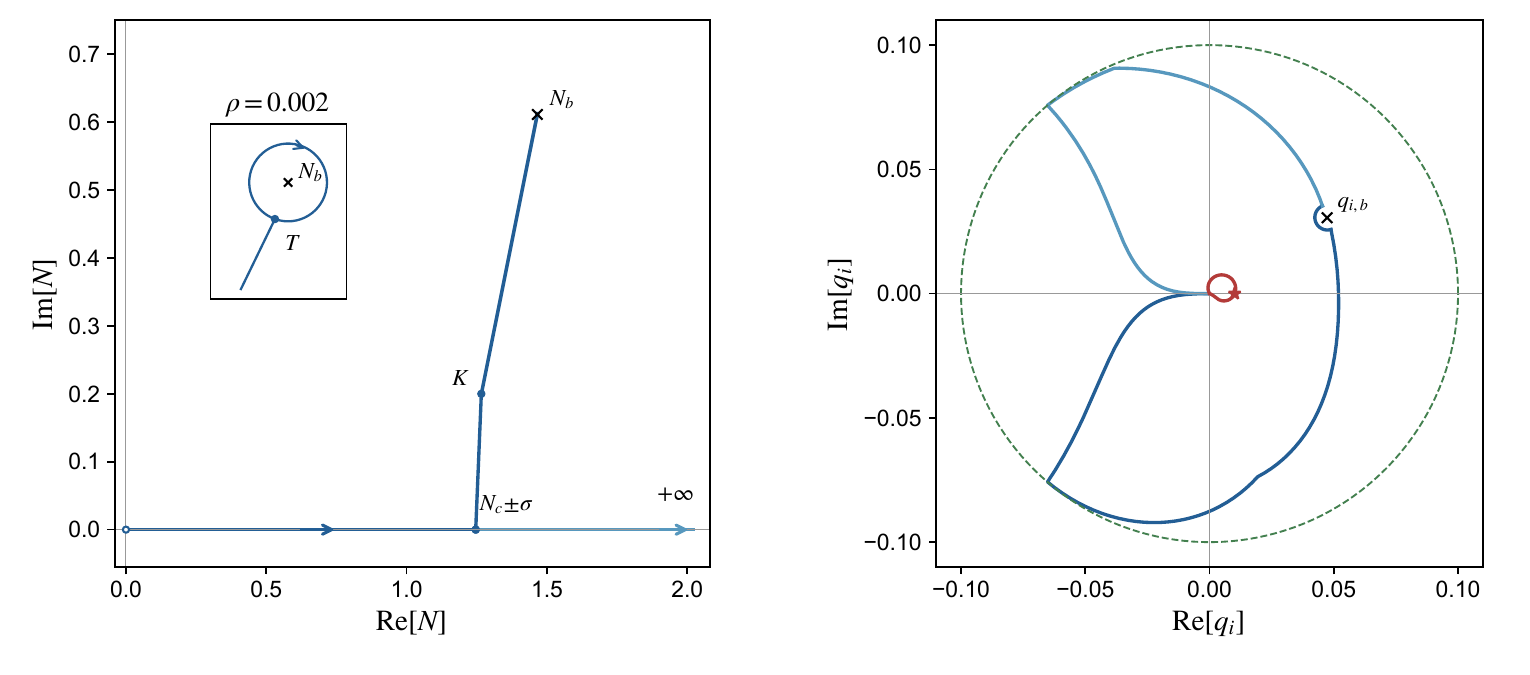}
 \caption{Positive-real-lapse portions joined by $\gamma_\rho$ (left) and their continued $q_i$ contour (blue, right), for $H=1$, $\epsilon=0.01$, and $q_1=2$.  The inset enlarges the clockwise circle of radius $\rho=0.002$ around $N_b$.  The open point and arrow indicate the limiting ends $N=0$ and $+\infty$.  The dark and light blue curves distinguish incoming and outgoing portions and the red curve shows the steepest-descent contour $D_1+D_2$.  The star marks the tunneling saddle and the cross marks $q_{i,b}$. The dashed green circle represents $|q_i|=R_\epsilon$.}
 \label{fig:real-lapse-connection}
\end{figure*}

The red contour $D_1+D_2$ in the right panel of Fig.~\ref{fig:real-lapse-connection} can be continuously deformed into the blue contour on this same branch, with the endpoint radii kept small and nonzero.  The deformed contour has two positive-real-lapse portions connected by $\gamma_\rho$.  We denote by $N_c$ the real lapse shared by two conjugate points on $|q_i|=R_\epsilon$ (see also Sec.~\ref{sec:circle-action}). In this example $N_c\simeq 1.25$.  Choosing $K=N_c+0.02+0.2i$, $\sigma=10^{-4}$, and $\rho=0.002$ with 
\begin{equation}
 T=N_b-\rho\frac{N_b-N_c}{|N_b-N_c|},
\end{equation}
we take
\begin{align}
 \gamma_\rho:\quad N_c-\sigma&\longrightarrow K\longrightarrow T,
 \nonumber\\
 &\overset{\text{one clockwise circuit}}{\longrightarrow}
 T\longrightarrow K\longrightarrow N_c+\sigma.
 \label{eq:real-lapse-connection}
\end{align}
The circuit is $N=N_b+\rho e^{i\theta}$, with $\theta$ decreasing from $\theta_0=\arg(T-N_b)$ to $\theta_0-2\pi$.  The points $K$ and $T$ only specify the path.  The part between $K$ and $T$ is traversed in both directions in the $N$ plane.  The return follows the other $q_i$ solution, so the two traversals must be kept separately in the integral.

The first real portion follows the short branch from small $N$ to $N_c-\sigma$; the second follows the long branch from $N_c+\sigma$ to large $N$.  The corresponding $q_i$ contour has $\max|q_i|\simeq0.09999<R_\epsilon$ and crosses the real axis at $q_i\simeq0.0519$, where the two formulas in App.~\ref{app:B} join.  We checked the same connection for $\rho=0.01,0.005,0.002,0.001$.  The endpoints of $\gamma_\rho$ and their continued actions agree in these checks.  The deformation can be defined by interpolating $\log q_i$ separately on the lower and upper portions, joining them on $q_* \equiv \epsilon / (H^2 q_1) <q_i<R_\epsilon$.  The radius then remains nonzero and smaller than $R_\epsilon$ throughout the interpolation.

This verifies a continuous deformation of the chosen classical contour at finite endpoint cutoffs.  Extending it to $N=0$ and $N=\infty$ also requires a convergence prescription at those ends and continuation of the fluctuation factor.

\subsection{Action difference at the same real lapse}
\label{sec:circle-action}

Here we give further details of the two solutions on the circle $\abs{q_i} = R_\epsilon$.  
For $q_i=R_\epsilon e^{i\theta}$, $0<\theta<\pi$, the partner is $q_i'=q_i^*$.  Let $\Delta$ denote the continued upper value minus the lower value at these two circle points, and put $p=\sin^2(\theta/2)$.  Writing
\begin{align}
 K(p)&=\int_0^{\pi/2}\frac{du}{\sqrt{1-p\sin^2u}},\nonumber\\
 E(p)&=\int_0^{\pi/2}\sqrt{1-p\sin^2u}\,du,
\end{align}
the integral between the partner branch points gives
\begin{align}
 \Delta N&=\frac{i\sqrt{R_\epsilon}}{H}\,[2E(p)-K(p)],\nonumber\\
 \Delta W&=\frac{8i}{3}HR_\epsilon^{3/2}
       [(1-p)K(p)+(2p-1)E(p)],\nonumber\\
 \Delta S_\epsilon^{\rm cl}&=6\pi^2[C\Delta N-\Delta W],
 \label{eq:circle-periods}
\end{align}
where $C=1-2H^2R_\epsilon\cos\theta$ and $W$ is the integral of $\sqrt{C-U(q)}\,dq$ in the action.  The orientation runs from the lower circle point to the upper one.  Returning on the opposite square-root sheet doubles these differences.  These are integrals between branch points, not residues of a pole.

The condition $\Delta N=0$ has the unique solution $K(p_*)=2E(p_*)$, with $p_*\simeq0.826$ and $\theta_*\simeq2.28$, independent of $H$, $\epsilon$, and $q_1$.  The two circle points therefore have the same real lapse $N_c$.  Changing $\operatorname{Im}q_i$ to $-\operatorname{Im}q_i$ from the lower point to the upper one gives
\begin{align}
 \Delta S_\epsilon^{\rm cl}
 &=-16\pi^2iHR_\epsilon^{3/2}E(p_*),\nonumber\\
 &=-\frac{16\pi^2iE(p_*)}{\sqrt H}\epsilon^{3/4}\ne0.
 \label{eq:circle-matched-lapse-action}
\end{align}

To compare both solutions at $N_c$, complete $\gamma_\rho$ with the short real segments $N_c\to N_c-\sigma$ and $N_c+\sigma\to N_c$.  Its continued $q_i$ endpoints are $R_\epsilon e^{-i\theta_*}$ and $R_\epsilon e^{i\theta_*}$, with $\theta_*\simeq2.28$.  At these conjugate circle points, $\operatorname{Im}q_i$ changes to $-\operatorname{Im}q_i$.  Their upper-minus-lower action difference is given in Eq.~\eqref{eq:circle-matched-lapse-action}.  For the plotted parameters, $\Delta S_\epsilon^{\rm cl}\simeq-5.80i$.  Omitting the detour and joining the two real-lapse solutions directly at $N_c$ would give this jump.  The action varies continuously along the detour, and its net change is $\Delta S_\epsilon^{\rm cl}$ when the short real segments are included.  The two circle points can thus be connected through the disk, although they cannot simply be identified.

In a real-lapse representation the connecting path $\gamma_\rho$ must be retained.  Although $\Delta S_\epsilon^{\rm cl}$ tends to zero as $\epsilon^{3/4}$ at fixed $H$, it does not determine the integral of $a_\epsilon\exp(iS_\epsilon^{\rm cl})\,dN$ along that path.  Whether its effect on the normalized wave function vanishes as $\epsilon\to0$ is left for future work.  This requires the fluctuation factor, phase, and coefficient in the original endpoint integral.  The classical connection alone does not imply an additional independent contribution to that integral.  Near $N_b$, the two endpoint solutions must be treated together as we discuss in App.~\ref{app:min-measure}.

\subsection{Comparison with the outer-turning-point saddle}
\label{sec:outer-region}

\begin{figure}[t]
    \centering
    \includegraphics[width=0.45\textwidth]{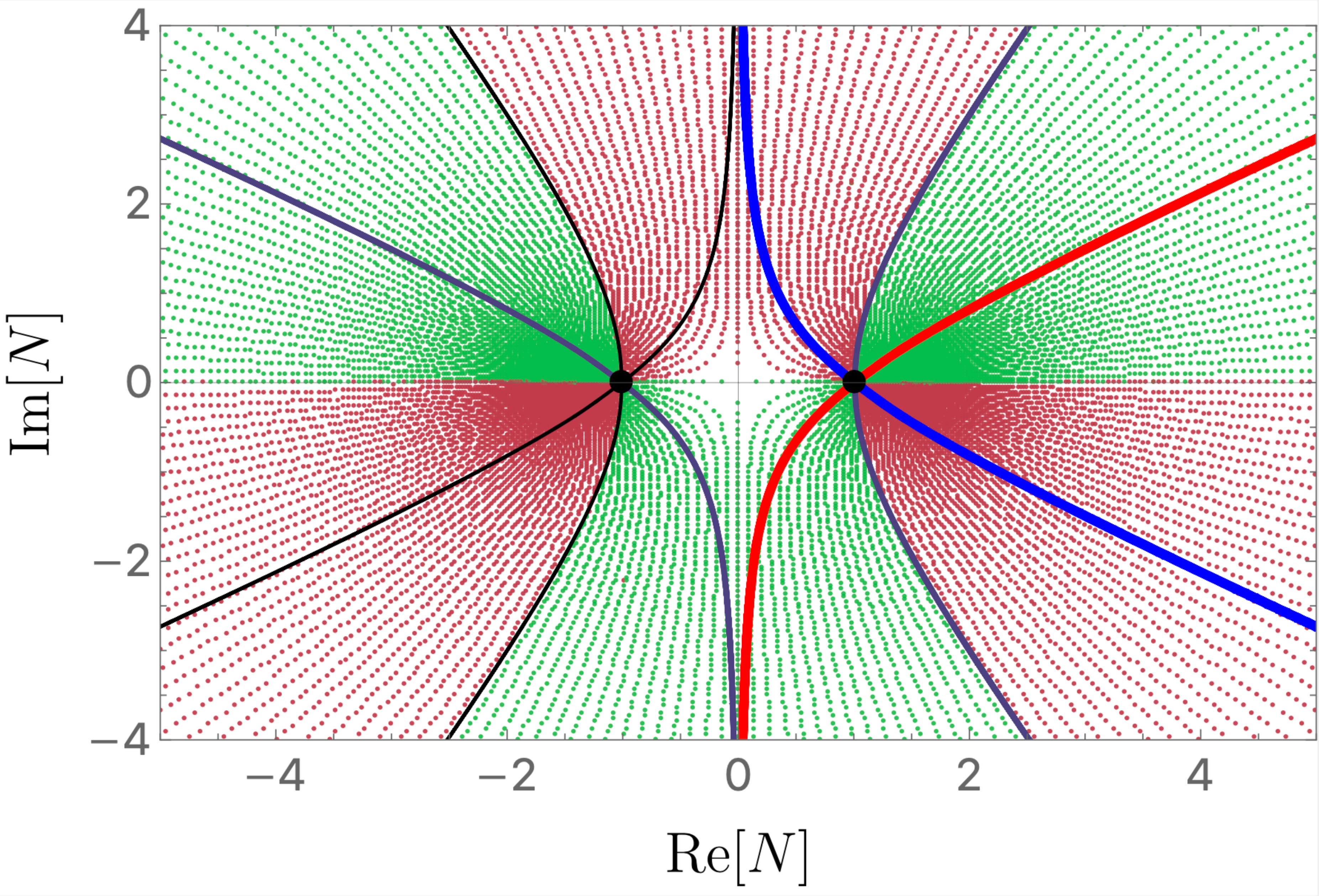}
    \vspace{0.4cm}
    \\
    \includegraphics[width=0.45\textwidth]{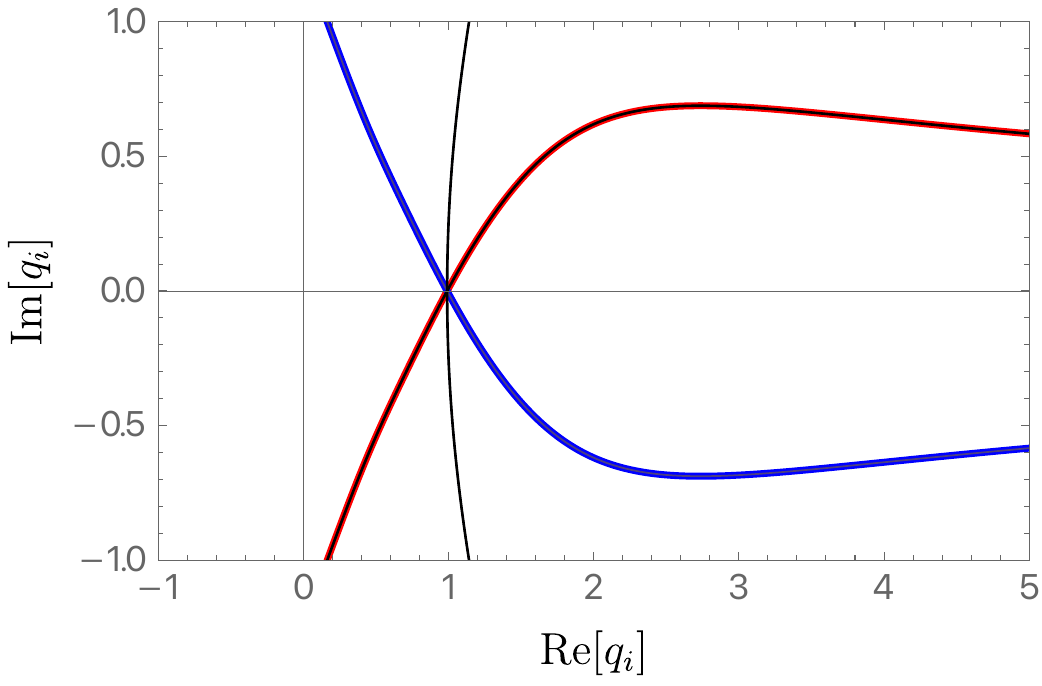}
    \caption{
    Complex lapse plane for classical solutions obeying the Neumann condition, obtained from the complementary region $|q_i|>R_\epsilon$, which includes the outer turning point $q_+$.
    The lower panel shows the corresponding contour in the $q_i$ plane.
    The additional real-lapse saddle in this region is omitted by the chosen small-throat contour.
}
    \label{fig:large-domain-plane}
\end{figure}

For comparison, 
we also plot the case with 
$|q_i| > R_\epsilon$ in Fig.~\ref{fig:large-domain-plane}.
We sample $q_i =r e^{i\theta}$ on a $200\times100$ grid, with $\ln r$ and $\theta$ uniformly spaced in the ranges $r \in (\sqrt{\epsilon}, 10^2)$ and $\theta \in (0,2\pi)$. 
The green and red dots represent sample points satisfying $\operatorname{Re}[iS_{\epsilon}^{\rm cl}(N)] < \operatorname{Re}[iS_{\epsilon}^{\rm cl}(N_{+,*}^{(1)})]$
and $\operatorname{Re}[iS_{\epsilon}^{\rm cl}(N)] > \operatorname{Re}[iS_{\epsilon}^{\rm cl}(N_{+,*}^{(1)})]$, respectively. 
The red and blue curves show the steepest-descent and steepest-ascent contours associated with the saddle $N_{+,*}^{(1)}$, while the black curves show the corresponding constant-phase contours for the other branches.  The gray curve denotes a contour of constant $\operatorname{Re}[iS_{\epsilon}^{\rm cl}] = \operatorname{Re}[iS_{\epsilon}^{\rm cl}(N_{+,*}^{(\pm1)})]$. 
Since this domain includes $q_+$, an additional unsuppressed saddle appears on the real lapse axis.
This saddle is the outer-turning-point saddle described by Eq.~\eqref{eq:Nstar_plus}. 
Its presence illustrates why the bare Neumann condition is not sufficient and why the choice of the small-throat cycle is an essential part of the prescription.

\section{Discussion and conclusions}
\label{sec:discussion}

We have proposed a small-throat prescription for the tunneling wave function
of a closed universe.  The prescription is motivated by the idea that universe
creation can be understood as a pinch-off, or decoupling, limit of a tunneling
geometry connected to another spacetime through a small throat.  In the
effective minisuperspace description used in this paper, the finite throat is
represented by the inner turning point generated by a small radiation
component.

The prescription has two essential ingredients.  The first is the Neumann
condition obtained by varying the initial endpoint at the saddle, $\dot q(0)=0$. 
The second is that the initial value $q_i=q(0)$ is restricted to a relative
cycle $D_\epsilon$ contained in the punctured disk, $0<|q_i|<R_\epsilon =\sqrt{\epsilon}/H$. 
The corresponding steepest-descent lapse contour passes through
the inner tunneling saddle 
and excludes the unsuppressed outer-turning-point saddle at $q_i=q_+$. 
We have not calculated all saddle contributions to the
original integral over the initial endpoint and real lapse.  Thus a
finite-throat-motivated contour can be chosen to give the tunneling saddle,
but the throat alone has not been shown to select that contour uniquely.
We assume that the integration contour is given by the steepest-descent lapse contour and consider the corresponding saddle action in the semiclassical approximation. 
The result established here is the limit of the leading saddle action and outgoing WKB phase. The validity of the semiclassical approximation in this limit requires further analysis.

The radiation component therefore plays a structural role.  It is not merely a
small perturbation of the pure de Sitter model.  At finite $\epsilon$, the
radiation term resolves the degenerate starting point into a finite inner
turning point and defines a nondegenerate saddle problem.

After this finite-$\epsilon$ problem is defined, we take the
zero-radiation limit.  In this limit,
\begin{equation}
    q_-\to0,
    \qquad
    q_+\to H^{-2},
    \qquad
    R_\epsilon\to0.
\end{equation}
The tunneling exponent and Lorentzian WKB phase then reduce to their pure de
Sitter values at fixed $q_1>H^{-2}$, reproducing the leading exponential factor of the standard tunneling wave function.  
It is important, however, that the limiting boundary condition
should not be confused with a boundary condition imposed directly after
setting $\epsilon=0$.  In the usual pure de Sitter minisuperspace problem with
$q(0)=0$, the tunneling saddle has a generally complex initial derivative. In
the conventions used here one may write, for the outgoing branch,
\begin{equation}
    \dot q(0)
    =
    \frac{2}{H^2}
    \left(
        1-i\sqrt{H^2q_1-1}
    \right).
\end{equation}
In the small-throat family, by contrast,
\begin{equation}
    \dot q(0)=0,
    \qquad
    q(0)=q_-=
    \frac{1-\sqrt{1-4H^2\epsilon}}{2H^2}
    \to0 .
\end{equation}
The two procedures lead to the same tunneling saddle action and outgoing
WKB phase in the $\epsilon\to0$ limit, but the finite-$\epsilon$ boundary value problems are
different.  This is the main point of the construction.

The present work is deliberately limited to the homogeneous minisuperspace
sector.  We have not derived the small-throat cycle from a complete two-sided
wormhole or parent-universe path integral.  In such a more complete
description, the radiation parameter $\epsilon$ should be replaced by the
matter content or charge that supports the finite throat, and the chosen
lapse contour, whose image is $D_\epsilon$, should arise from the full path integral of the
two-sided geometry.  It would be interesting to determine whether this can be
done explicitly in known Euclidean wormhole or throat geometries, and whether
the same inversion scale $R_\epsilon=\sqrt{q_-q_+}$ has a direct geometric
interpretation there.

Finally, we comment on 
the problem of unsuppressed or inverse-Gaussian
perturbations in Lorentzian quantum cosmology emphasized in
Refs.~\cite{Feldbrugge:2017fcc,Feldbrugge:2017mbc,Feldbrugge:2018gin,
Bojowald:2018gdt,Wang:2019spw,Bojowald:2020kob,Matsui:2022lfj,
Matsui:2024bfn}.  Possible responses and modified prescriptions, including
tunneling boundary conditions, Neumann/Robin boundary conditions, and initial-state
regularizations, have been discussed in
Refs.~\cite{Vilenkin:2018dch,Vilenkin:2018oja,DiTucci:2019dji,
DiTucci:2019bui,DiTucci:2020weq,Yamada:2025rld,Ailiga:2023wzl,
Ailiga:2024mmt,Ailiga:2024wdx,Mallik:2026iai}.  
In the small-throat prescription considered in this paper, the starting point is
not the singular zero-size geometry itself, but a finite-radiation tunneling
problem with a nonzero inner turning point.  
At any finite value of
$\epsilon$, the background is therefore an ordinary tunneling problem from a
small but finite universe.  
This motivates considering regular tunneling boundary conditions for the fluctuation modes at the finite throat as a possible way to address the unsuppressed-perturbation problem.
However, it remains to be determined whether these boundary conditions give suppressed fluctuations, whether any such suppression survives the limit $\epsilon\to0$, and whether the resulting state approaches the tunneling initial condition considered in earlier analyses~\cite{Vilenkin:2018dch,Vilenkin:2018oja,Yamada:2025rld} in this limit.
An explicit perturbation calculation is left for future work.

\section*{Acknowledgements}
This work was supported by the Japan Society for the Promotion of Science (JSPS) through Grants-in-Aid for Scientific Research (KAKENHI), Grant No. JP23K13092.

\appendix

\section{Integration measure and the semiclassical prefactor}
\label{app:min-measure}
In the main part of this paper, we calculate the saddle action.  To obtain the full
wave function, one also needs an integration measure, the Gaussian
fluctuation factor, and a normalization.  Here we state the measure used
for the minisuperspace model and examine the local fluctuation factor.

We use the canonical phase-space measure and the normalization
$\langle q|q'\rangle=\delta(q-q')$, so the initial-endpoint measure is $dq$.
In the constant-lapse gauge we take the lapse measure to be $dN$, up to
an overall normalization.

We denote by $q(t;q_i,N)$ the classical solution with
$q(0)=q_i$ and $\dot q(0)=0$ for a given $N$.
We define $F(q_i,N)=q(1;q_i,N)-q_1$ and
$w_f=\dot q(1)/(2N)$, keeping the sign of the final velocity.
The final boundary condition is $F(q_i,N)=0$.
On the chosen branch, we write its solution as $N=N(q_i)$.

To relate the fluctuation determinant to the classical solution,
let us define $y(t)=\partial q(t;q_i,N)/\partial q_i$ at fixed $N$.
It satisfies the linearized equation with
$y(0)=1$ and $\dot y(0)=0$.
For a Neumann condition initially and a Dirichlet condition
finally, the fluctuation determinant divided by the free
determinant is $y(1)$, since the corresponding free solution
is $y_0(t)=1$. Thus,
\begin{equation}
 J=y(1)=\left.\frac{\partial F}{\partial q_i}\right|_N.
\end{equation}
Differentiating $F(q_i,N(q_i))=0$ gives
\begin{equation}
 J=-2w_fN'(q_i),
\end{equation}
where we used
$\partial F/\partial N=\dot q(1)/N=2w_f$.
We therefore obtain
\begin{equation}
 dN\,J^{-1/2}
 =\frac{N'(q_i)\,dq_i}{\sqrt{-2w_fN'(q_i)}}.
\end{equation}
For constant initial weight, $J^{-1/2}$ is the Gaussian
fluctuation factor near a regular classical solution,
with the square-root phase followed along the contour.

At $q_{i,b}$, where $N=N_b$, $N'(q_{i,b})=0$ gives $J=0$.
Near this point,
\begin{equation}
 N-N_b\simeq
 \frac12 N''(q_{i,b})(q_i-q_{i,b})^2,
\end{equation}
so two nearby initial-endpoint saddles merge at $N_b$.
The Gaussian approximation for either saddle separately
then fails.
The separate Gaussian expression behaves as
$J^{-1/2}\propto(N-N_b)^{-1/4}$ and, for finite classical
action, would give an integral of $O(\rho^{3/4})$ around
a shrinking circle of radius $\rho$.
This estimate concerns only the circle, not the full
connecting path $\gamma_\rho$.

The two saddles must instead be treated together,
keeping the cubic term in $S_D(q_i,N;q_1)$.
Which combination contributes depends on the original
endpoint contour.
Consequently, the branch connection in the classical
solutions does not by itself establish an additional
independent physical contribution to
Eq.~\eqref{eq:path_integral}.
This joint treatment is left for future work.

At fixed $\epsilon>0$ and $q_i\to0$, the measures including this
fluctuation factor are $O(1)dq_i$ on the short branch and
$O(q_i^{-1/2})dq_i$ on the long branch.  These powers do not overcome the
exponential decay strictly inside the angular ranges specified in
Sec.~\ref{sec:branch-continuation}. However, this checks convergence near the ends only. 
The continuation of the full fluctuation factor and the
$\epsilon\to0$ limit of the normalized wave function require further work.

\section{Alternative method for contour choice}
\label{app:B}

In Sec.~\ref{sec:branch-continuation}, we explain how to calculate the contour integral in our numerical implementation. 
When we follow the same tunneling solution as $q_i$ varies, we instead use an alternative method explained in this Appendix.

Recall that the other branch point is $q_i'=\epsilon/(H^2q_i)$. It coincides with the final endpoint $q_1$ when $q_i=q_*$, where we define $q_*\equiv\epsilon/(H^2q_1)<q_-$.
Inside $|q_i|<R_\epsilon$, we use a cut along the real segment
$-R_\epsilon<q_i\leq q_*$.  The two sides of this cut retain different
values of the continued integrals.
We put
$q(u)=q_i+(q_1-q_i)u^2$, $0\leq u\leq1$, and define the direct integrals
\begin{align*}
 N_d(q_i)&=\frac{\sqrt{q_1-q_i}}{H}\int_0^1
       \frac{\sqrt{q(u)}}{\sqrt{q(u)-q_i'}}\,du,\\
 S_d(q_i)&=6\pi^2U(q_i)N_d(q_i)\nonumber\\
 &\quad-12\pi^2H(q_1-q_i)^{3/2}\int_0^1u^2
       \frac{\sqrt{q(u)-q_i'}}{\sqrt{q(u)}}\,du.
\end{align*}
We use the principal square roots in these direct integrals.  They vary
analytically within each open half of the $q_i$ plane.  
Writing $(N,S)$ for
the lapse and action on the tunneling branch, the required continuation is
\begin{align}
 &(N,S)(q_i)\nonumber\\
 &\quad=
 \begin{cases}
 (N_d,S_d)(q_i),&\Im q_i<0,\\
 2(N_d,S_d)(q_i')-(N_d,S_d)(q_i),&\Im q_i>0.
 \end{cases}
 \label{eq:min-branch-atlas}
\end{align}
In evaluating $(N_d,S_d)(q_i')$, the starting point is $q_i'$ and its
partner branch point is $q_i$.  The action uses $U(q_i')=U(q_i)$.

To check if these expressions join across the real axis, take
$q_*<q_i<R_\epsilon$ and write
\begin{align*}
 I&=\int_{q_i}^{q_i'}\frac{dq}{\sqrt{U(q)-U(q_i)}},\nonumber\\
 L&=\int_{q_i'}^{q_1}\frac{dq}{\sqrt{U(q_i)-U(q)}}.
\end{align*}
Then $N_d(q_i\pm i0)=(L\mp iI)/2$ and $N_d(q_i'-i0)=L/2$.
The expression for $\Im q_i>0$ in Eq.~\eqref{eq:min-branch-atlas} therefore joins
the lower one with the same value $(L+iI)/2$.  Splitting the action integral
along the same contours gives the corresponding identity for $S$.
Thus the solution can be followed continuously
through the interval $q_*<q_i<R_\epsilon$.

For $0<q_i<q_*$, however, the two sides of the cut give different values.
They lead to the short and long branches described in
Eqs.~\eqref{eq:qi-origin-short} and \eqref{eq:qi-origin-asymptotic}.

\bibliography{reference}

\end{document}